# Supermassive black hole mass measurements for NGC 1300 and NGC 2748 based on HST emission-line gas kinematics

J. W. Atkinson,[1*] J. L. Collett,[1] A. Marconi,[2] D. J. Axon,[1,3] A. Alonso-Herrero,[4] D. Batcheldor,[1,3] J. J. Binney,[5] A. Capetti,[6] C. M. Carollo,[7] L. Dressel,[8] H. Ford,[9] J. Gerssen,[10] M. A. Hughes,[1] D. Macchetto,[11] W. Maciejewski,[12] M. R. Merrifield,[13] C. Scarlata,[7] W. Sparks,[8] M. Stiavelli,[8] Z. Tsvetanov,[9] and R. P. van der Marel[8]

[1] *Centre for Astrophysics Research, STRI, University of Hertfordshire, Hatfield, Hertfordshire, AL10 9AB, UK*
[2] *INAF, Osservatorio Astrofisico di Arcetri, Largo Enrico Fermi 5, I-50125 Florence, Italy*
[3] *Department of Physics, Rochester Institute of Technology, 84 Lomb Memorial Drive, Rochester, NY 14623-5603, USA*
[4] *Departamento de Astrofísica Molecular e Infrarroja, IEM, Consejo Superior de Investigaciones Científicas, Serrano 113b, 28006 Madrid, Spain*
[5] *University of Oxford, Theoretical Physics, Keble Road, Oxford, OX1 3NP, UK*
[6] *INAF, Osservatorio Astronomico di Torino, Strada Osservatorio 20, I-10025 Pino Torinese, Torino, Italy*
[7] *Institute of Astronomy, Physics Department, ETH, Zurich, Switzerland*
[8] *Space Telescope Science Institute, 3700 San Martin Drive, Baltimore, MD21218, US*
[9] *Department of Physics and Astronomy, John Hopkins University, 3400 North Charles Street, Baltimore, MD21218, USA*
[10] *Dept. of Physics, Rochester Building, Science Laboratories, South Road, Durham, DH1 3LE, UK*
[11] *Space Telescope Division of ESA, Space Telescope Science Institute, 3700 San Martin Drive, Baltimore, MD21218, USA*
[12] *Obserwatorium Astonomiczne Uniwersytetu Jagiellonskiego, Poland.*
[13] *School of Physics and Astronomy, University of Nottingham, Nottingham, NG7 2RD, UK*

**ABSTRACT**

We present Space Telescope Imaging Spectrograph emission-line spectra of the central regions of the spiral galaxies NGC 1300 and NGC 2748. From the derived kinematics of the nuclear gas we have found evidence for central supermassive black holes in both galaxies. The estimated mass of the black hole in NGC 1300 is $\left(6.6^{+6.3}_{-3.2}\right) \times 10^7\ M_\odot$ and in NGC 2748 is $\left(4.4^{+3.5}_{-3.6}\right) \times 10^7$ (both at the 95% confidence level). These two black hole mass estimates contribute to the poorly sampled low-mass end of the nuclear black hole mass spectrum.

## 1. INTRODUCTION

The demography of supermassive black holes is an important element in understanding the many aspects of galaxy formation and evolution. Knowledge of the distribution function of black holes is fundamental in the formulation and testing of theories of black hole and galaxy formation. Initial studies concentrated primarily on early type systems (e.g. Macchetto et al. 1997; Magorrian et al. 1998); however, attention has recently been profitably given to spirals (e.g. Barth et al. 2001; Sarzi et al. 2001). Nevertheless, there is a limited number of nuclear black hole mass estimates and weak coverage of the mass spectrum, particularly at the low-mass end.

* email: jonnya@star.herts.ac.uk

The low-mass end of the mass spectrum is of particular importance for a number of reasons. Kormendy & Richstone (1995) showed, using early estimates for eight black hole candidates, that the black hole mass is correlated with the luminosity (or mass) of the host bulge. Magorrian et al. (1998) later strengthened this idea with a sample of 36 galaxies. More recently, both Ferrarase & Merritt (2000) and Gebhardt et al. (2000) have independently shown that a tighter correlation exists between hole mass and the velocity dispersion of the host bulge - although some authors have derived equally tight correlations based on luminosity (e.g. McLure & Dunlop 2002; Marconi & Hunt 2003). If these correlations are indeed intrinsic, they are of paramount importance in the sense that they can be used to estimate black hole masses at large distances where the resolution is too low for the practical use of more direct methods. The bulk of the black hole mass estimates are in the range of $10^7$ to $10^9$ $M_\odot$. Schödel et al. (2002) showed the mass of the Galactic black hole is $(3.7 \pm 1.5) \times 10^6$ $M_\odot$, using a combination of high-resolution imaging and spectroscopy, obtained over a period of ~ 10 years. Even this beautiful paper can only constrain the nearest nuclear black hole mass to ~ 40%. Clearly, there are nuclear black holes with masses that fall in the low-mass end of the spectrum, and the various correlations predict that the lower mass spheroids – spiral bulges – might harbour these low mass nuclear black holes. It is important to add to the current database of black hole mass estimates, regardless of the validity and tightness of these correlations.

We use a volume limited (recession velocity < 2000 km s$^{-1}$) sample of 54 galaxies that are taken from a larger sample of 128 galaxies all classified Sb, SBb, Sc or SBc for which we have ground-based H$\alpha$ and [NII] rotation curves. We used the Space Telescope Imaging Spectrograph (*STIS*) on the Hubble Space Telescope (*HST*) to obtain emission-line long-slit spectra for this sample. The spectra have been previously illustrated in Hughes et al. (2003) and a photometric analysis of the *STIS* acquisition images was presented by Scarlata et al. (2004). Interpretation and modelling of the spectra is far from straightforward. In some cases, the rotation curves of the galaxies are too poorly sampled or the emission-line gas delineates highly irregular velocity fields. These cases cannot be meaningfully analysed using a model in which the gas rotates solely in a coplanar axisymmetric disc. Nevertheless, there are galaxies in the sample that exhibit rotation curves to which such a model can be successfully applied. Two of these galaxies, NGC 1300 and NGC 2748, are the subject of this paper.

Table 1. Individual root-names for each spectrum file, exposure time, slit position and binning strategy, for each galaxy

| Galaxy | Data-set Name | Exposure Time (s) | Slit Position | On-chip Binning |
|---|---|---|---|---|
| NGC 1300 | o5h707010 | 240 | OFF1 | 2×2 |
|  | o5h707020 | 240 | OFF1 | 2×2 |
|  | o5h707030 | 360 | NUC | 1×1 |
|  | o5h707040 | 380 | NUC | 1×1 |
|  | o5h707050 | 240 | OFF2 | 2×2 |
|  | o5h707060 | 288 | OFF2 | 2×2 |
| NGC 2748 | o5h709010 | 288 | OFF1 | 2×2 |
|  | o5h709020 | 288 | OFF1 | 2×2 |
|  | o5h709030 | 432 | NUC | 1×1 |
|  | o5h709040 | 432 | NUC | 1×1 |
|  | o5h709050 | 288 | OFF2 | 2×2 |
|  | o5h709060 | 288 | OFF2 | 2×2 |

The first galaxy to be kinematically modelled was NGC 4041 (Marconi et al. 2003) where we found evidence of a black hole of mass ~ $10^7$ $M_\odot$ based on the standard assumptions of a nuclear disc of emission-line gas, coplanar with the galaxy. However, we also found that the central disc is blueshifted with respect to the extended disc suggesting that the two components may be kinematically decoupled. Consequently, in allowing the standard assumptions to be relaxed (e.g. allowing the mass-to-light ratio of the nuclear component to be different from that of the extended component), we were only able to set an upper limit to the mass of the black hole of ~ $6 \times 10^6$ $M_\odot$. Marconi et al. (2003) establishes the methodology that we use in investigating galaxies within the sample and we follow this approach in modelling NGC 1300 and NGC 2748.

## 2. OBSERVATIONS AND DATA REDUCTION

NGC 1300 is a striking galaxy with a strong well-defined kpc-scale bar; it does not exhibit nuclear activity and is classified SBbc (de Vaucouleurs et al. 1991). The classical bar structure lends itself to dynamical modelling: the HI velocity field within the bar and spiral arms has been modelled with tilted rings (Lindblad et al. 1997) and the large-scale gravitational potential inferred from a mass model based on the decomposition of the galaxy into individual structural components – the bulge, bar, disc and lens (Aguerri et al. 2001). Martini et al. (2003) note that the nuclear dust lanes connect to those of the leading edge of the bar. This suggests that the large and small-scale gas flows are intimately linked. LEDA[1] quotes the average radial velocity (by radio measurements) of NGC 1300 as 1575 ± 18 km s$^{-1}$, which becomes 1409 km s$^{-1}$ after correction for Local Group infall into Virgo. We thus adopt a distance of approximately 18.8 Mpc, at which distance 1″ corresponds to ~ 91 pc (based on an $H_0$ of 75 km s$^{-1}$ Mpc$^{-1}$). The position angle of the large-scale disc is ~ 106° and its inclination is ~ 49° (LEDA), both of which we adopt throughout this analysis. These values are consistent with those derived by Elmegreen et al. (1996) and those published in England (1989); England derived the position angle to be ~ 95°, offset to our adopted value by only 10°.

NGC 2748 is an unbarred galaxy classified as SAbc (de Vaucouleurs et al. 1991). Its radial velocity (by radio measurement) is ~ 1476 ± 8 km s$^{-1}$: this becomes ~ 1741 km s$^{-1}$ when corrected for the Local Group infall. For $H_0 = 75$ km s$^{-1}$ Mpc$^{-1}$, this yields a distance of 23.2 Mpc, which means that one arcsecond corresponds to a linear distance of approximately 112 pc. The galaxy is reasonably highly inclined; LEDA reports the inclination to be ~ 73°. It might be thought that an edge-on disc provides the optimal orientation to maximise the amplitude of the observed rotation curve and indeed, this will be the case for an infinitesimally thin slit. For a finite slit width (or indeed a disc of non-negligible thickness), the velocities sampled at some point along the slit originate from a range of radii on the disc. The corresponding average velocity can, for certain rotation curves, decrease. An obvious exception is that of a disc in solid body rotation where the radial velocity across the slit is constant. In this case the edge-on disc yields the maximum amplitude in the rotation curve. Given the typical rotation curves we observe in the centres of spirals, we expect the inclination of NGC 2748 to yield a near-maximum amplitude rotation curve.

Our observational strategy consists of obtaining an acquisition image and then two spectra at three parallel slit positions for each target. Full details of this strategy, the instrumental set-up and data reduction procedures can be found in Marconi et al. (2003). Specific details of the instrumental set-up for both NGC 1300 and NGC 2748 are given in Table 1. The position angles of the slits for NGC 1300 and NGC 2748 are 110° and 38° east of north, respectively. The spectra of NGC 1300 were obtained at a position angle offset of only 4° with respect to the kinematic line of nodes (i.e. the kinematic line of nodes of the galaxy). Regarding NGC 2748, we obtained all spectra with the slits placed along the major axis of the galaxy.

---
[1] Available at http://leda.univ-lyon1.fr

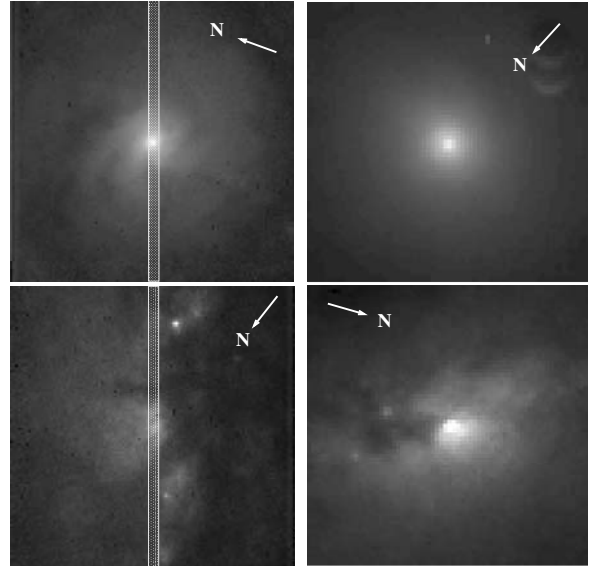

Figure 1. Left: *STIS* F28X50LP filter acquisition images (NGC 1300 is top, NGC 2748 lower). Right: *NICMOS* F160W filter images. The F160W images have been cropped to match the STIS acquisition images in size (5″×5″). The NUC slits are indicated in the two acquisition images: the OFF1/2 slits are the same size and are located immediately adjacent to the NUC slits.

For our two galaxies we have retrieved *NICMOS* images from the archive. The images were obtained using the F160W filter (approximately H-band), have a pixel scale of 0.076″, and are of dimensions 19.2″ × 19.2″. For details of the data reduction techniques used on these images, consult Hughes et al. (2003).

We have also made use of the WFPC2 Associations archive to retrieve 'science-ready' F606W filter images. For details of the reductions and registration of the *WFPC2* images, the reader should consult the webpage[2].

We display rotation curves derived from ground-based observations obtained using IPCS on the Isaac Newton Telescope on La Palma (Vila Costas 1991). NGC 1300 was observed on 1985 October 16 with an exposure time of 1000 seconds and NGC 2748 was observed on 1987 June 5 with an integration time of 1200 seconds. No analysis is made using these rotation curves – we simply compare them with our high-resolution rotation curves.

## 3. MODELLING TECHNIQUES

In modelling the observed rotation curves we use the standard techniques applied when studying emission-line gas kinematics (e.g. Macchetto et al. 1997; van der Marel & van den Bosch 1998; Barth et al. 2001; Marconi et al. 2003).

To account for the contribution to the gravitational potential of the stellar bulge, we model the surface brightness profiles using the radial density profile model described by van der Marel & van den Bosch (1998). These models (hereafter, V98 components) are defined as follows

$$\rho(m) = \rho_0 \left(\frac{m}{r_b}\right)^{-\alpha} \left(1 + \left(\frac{m}{r_b}\right)^2\right)^{-\beta}$$

where $\rho_0$ is a scale density and $r_b$ is a scale length. We shall not describe the mathematical details of how such a three-dimensional density model is converted into a one-dimensional surface brightness profile here, but refer the reader to Appendix A. of Marconi et al. (2003).

We use a simple razor-thin disc for the geometry of the ionised gas. The gas is assumed to travel along circular orbits in a disc, which itself lies in the principal plane of the potential. We assume that the motion is entirely due to the combined gravitational potential of the stellar component and any putative black hole; thus we neglect pressure effects and turbulence in our model.

The velocity information is essentially contained in the full profile of the emission lines; from this profile we infer the flux of the emission line, the velocity centroid and the velocity dispersion. The emission lines observed in our spectra are well approximate by single Gaussians.

[2] http://archive.stsci.edu/hst/wfpc2/

Therefore, the complete velocity profile is not required for the fitting routines. Instead, we use the first three moments of the flux-weighted velocity distribution. Again, the mathematical model used to derive these moments is to be found in Appendix B. of Marconi et al. (2003); see equations B10 to B12. The model takes into account the PSF of the telescope using images generated by TINYTIM (Krist & Hook 1999). We also model the finite width of the slit (see Maciejewski & Binney 2001 for a detailed discussion) and the finite size of the detector pixels.

Barth et al. (2001) showed that the large-scale emission-line flux distribution is an important element in modelling the micro-structure in the rotation curves in order to reduce the overall $\chi^2$ of the best-fitting model. The total emission-line flux at a point $x$, $y$ is represented in our model by the sum of the flux given by:

$$I(x, y) = \sum_i I_{0i} f_i \left(\frac{r_i}{r_{0i}}\right)$$

where $f_i$ is the chosen circularly symmetric function (e.g. a Gaussian or an exponential), $r_i$ is the radius, $r_{0i}$ is the characteristic scale radius of the function and $I_0$ is the weighting of the function.

There is no reason that structures observed in the emission-line flux should be circularly symmetric or be as smooth and continuous as our model suggests. Nevertheless, it is comforting that the rotation curves derived by Barth et al. (2001) using the emission-line images are similar to those derived when using an approximation similar to ours. There are small differences between both sets: those derived using the emission-line images reproduce some of the micro-structure in the observed rotation curves (as previously mentioned) whereas the rotation curves derived using the approximation to the emission-line flux distribution are generally smoother.

All of our models are optimised by $\chi^2$ minimisation using the downhill simplex routine of Press et al. (1992).

## 4. RESULTS

### 4.1 Bulge Density Models

The left panels Figure 1 show the acquisition images obtained for NGC 1300 (top) and NGC 2748 (bottom). In the case of NGC 1300 the telescope appears to have centred correctly; however, the acquisition image of NGC 2748 shows that the centre may well be obscured by dust. The model might return a large impact parameter if the telescope is indeed not centred on the kinematic centre. The F160W image of NGC 1300 is smooth and shows little evidence of dust. We also note that an 'artefact' is apparent to the west of the nucleus; this feature was masked out when fitting the ellipses. The dust lane visible in the acquisition image of NGC 2748 is also visible in the corresponding F160W image, but is less striking.

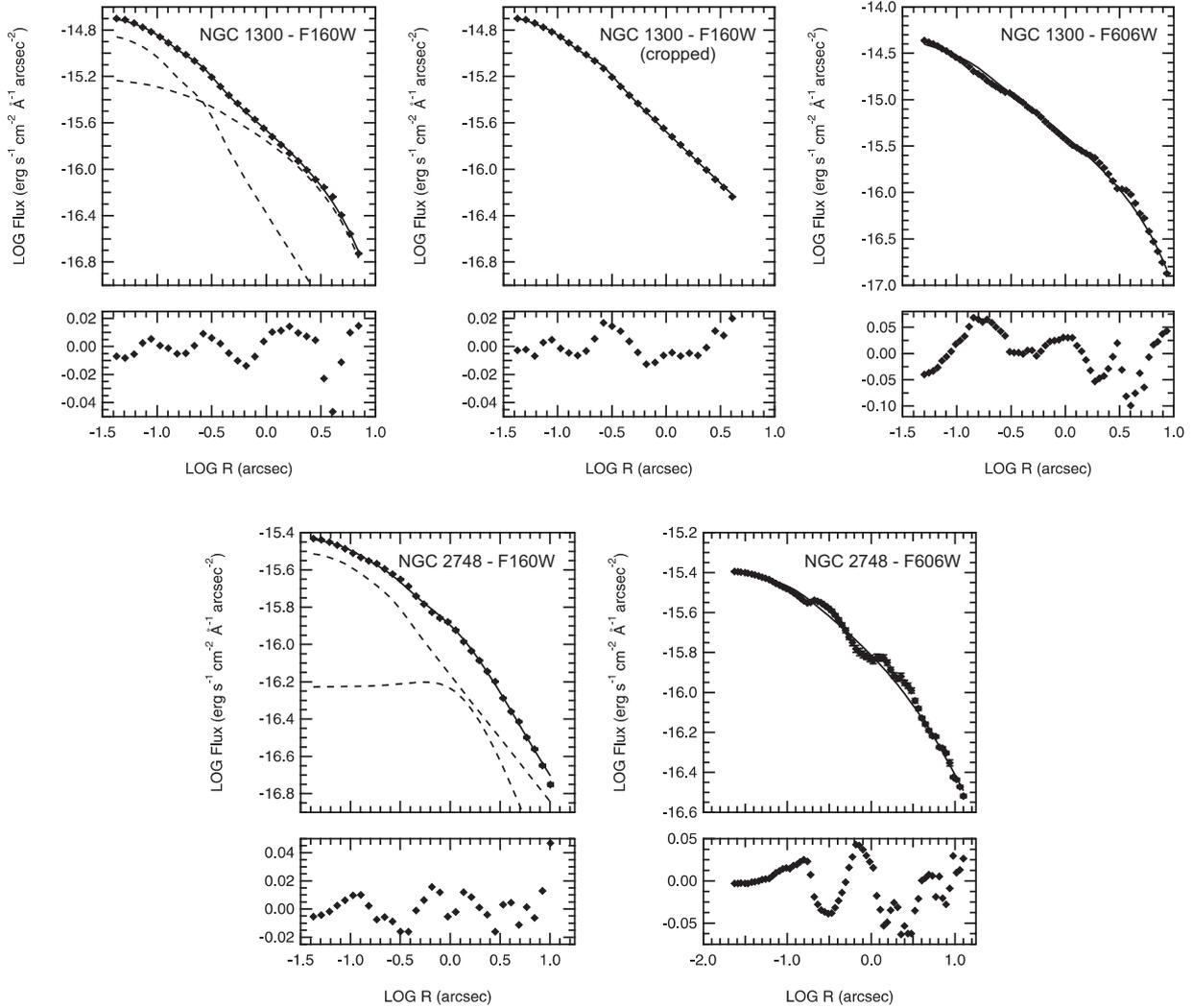

Figure 2. Bulge models (solid lines) overlaid on the surface brightness profiles of NGC 1300 and NGC 2748. The image from which each surface brightness profile is derived is labelled on each plot. The dashed lines show the individual V98 components. The lower panel shows residuals from the fit.

We derived surface brightness profiles from both the F160W and F606W images by fitting ellipses to the isophotes. The F160W images are displayed in the right panels of Figure 1, where they have been cropped in size to match the acquisition images. The F160W images are much less affected by dust than are optical images: one magnitude of extinction in the $V$ band corresponds to only ∼ 0.18 magnitudes in the $H$ band. Consequently, we consider the radial surface brightness profiles derived from the F160W images to be the best choice when approximating the bulge.

The F606W filter is a broadband filter centred on 5767Å and has a full width at half maximum (FWHM) of 1579Å, encompassing much of the Johnson $V$ and Cousins $R$ bands. To convert to the Cousins $R$ band we used SYNPHOT in IRAF based on bracketing blackbody temperatures of 4000 and 5000 K, resulting in an average correction of -0.2 magnitudes. This falls neatly in the range derived by Marconi et al. (2003) derived using galaxy and stellar spectra (K0V, Sb: ~ 0.4 and A0V, Sc: ~ 0.05) templates. There are techniques for mapping extinction but given that they are not particularly accurate we have chosen to approach the problem from a different perspective. Given that the F606W images will be significantly more affected by dust than the F160W images, we should determine different values for the black hole mass for bulge models based on different photometry if dust is problematic.

Table 2. Optimised parameters for the stellar potential in each model. $\rho_0$ is a scale density (assuming a mass-to-light ratio of unity), $r_b$ is the scale radius, $i$ is the assumed inclination, $q$ is the intrinsic axial ratio, and $\alpha$ and $\beta$ are the slopes of the profile.

| Galaxy | Band | $q$ | $i$ (°) | $\rho_0$ ($M_\odot$ pc$^{-3}$) | $r_b$ (arcsec) | $\alpha$ | $\beta$ | $\rho_0$ ($M_\odot$ pc$^{-3}$) | $r_b$ (arcsec) | $\alpha$ | $\beta$ | $\chi^2_{red}$ |
|---|---|---|---|---|---|---|---|---|---|---|---|---|
| NGC 1300 | F160W | 1 | N/A | $1.37 \times 10^4$ | 0.107 | 1.99 | 0.25 | 1.026 | 14.23 | 1.37 | 3.71 | 72.4 |
|  | F160W | 1 | N/A | $6.45 \times 10^4$ | 0.023 | 3.62 | -0.9 | - | - | - | - | 49.4 |
|  | F160W | 0.1 | 49 | $9.82 \times 10^3$ | 0.173 | 2.30 | -0.2 | - | - | - | - | 31.3 |
|  | F606W | 1 | N/A | 0.137 | 14.74 | 1.80 | 3.15 | - | - | - | - | 106.4 |
| NGC 2748 | F160W | 1 | N/A | $1.9 \times 10^3$ | 0.046 | 0.99 | 0.34 | 31.44 | 1.03 | -1.55 | 1.96 | 14.6 |
|  | F160W | 0.1 | 73 | $1.7 \times 10^3$ | 0.052 | 0.94 | 0.38 | 99.39 | 1.00 | -1.49 | 1.87 | 14.3 |
|  | F606W | 1 | N/A | 2.21 | 5.36 | 1.23 | 0.31 | - | - | - | - | 14.1 |

*NGC 1300*

The derived surface brightness profiles for NGC 1300 are displayed in the top panels of Figure 2. We have fitted two models to the profile derived from the F160W image; the first uses two V98 components (both spherical) to model the complete surface brightness profile. The surface brightness profile can be adequately reproduced up to a radius of approximately 4″ using a single V98 model. However, at a radius of ~ 4″ the surface brightness profile drops sharply over a region of about 6″ before levelling out. It is not clear whether this is a consequence of the galaxy being off-centre in the image, in which case the outer isophotes are only partially sampled, or something physical. Another contribution to the drop in the profile may be uncertainties in background subtraction. The limited sampling of the isophotes may be the cause of the steep 'tail' of the surface brightness profile at r > 4″. To investigate this we further trimmed the image so as to decrease the sampling of the isophotes and refitted ellipses. The trimming made little difference to the derived surface brightness profile so we might expect the 'tail' to be real. While this is an inconclusive test, it suggests it is indeed worth attempting to fit the complete profile with two V98 components as well as the truncated profile, with a single component. For completeness, we have fitted an additional model with a single V98 component to the truncated surface brightness profile where the mass distribution has an intrinsic axial ratio of 0.1. This model represents an extreme 'disc-like' case; the model fit is visually indistinguishable from that of the spherical one-component fit. Finally we have fitted a single V98 component to the surface brightness profile derived from the F606W image.

*NGC 2748*

The F160W image of NGC 2748 exhibits a dust feature southeast of the nucleus. We have derived surface brightness profiles with, and without, masks of the obvious dust features but found the difference between the two resultant profiles to be negligible. The resultant profile is smooth and does not appear to exhibit any strong shoulders or ripples. We first tried to model the surface brightness profile using a single V98 component but found that the profile was not satisfactorily reproduced at a radius greater than approximately 1″. Two V98 components match the surface brightness profile much better. The average axial ratio of the isophotes is ~ 0.8, which might be worrisome given our use of a spherical model for the bulge. To investigate this we fitted a second pair of V98 models with an observed axial ratio of 0.8. The resultant circular velocity profiles are almost identical to those of the spherical model: thus we are justified in choosing the spherical model to represent the bulge for simplicity. A second model in which the extended V98 component of the density model has a 'disc-like' axial ratio was also fitted to the surface brightness profile.

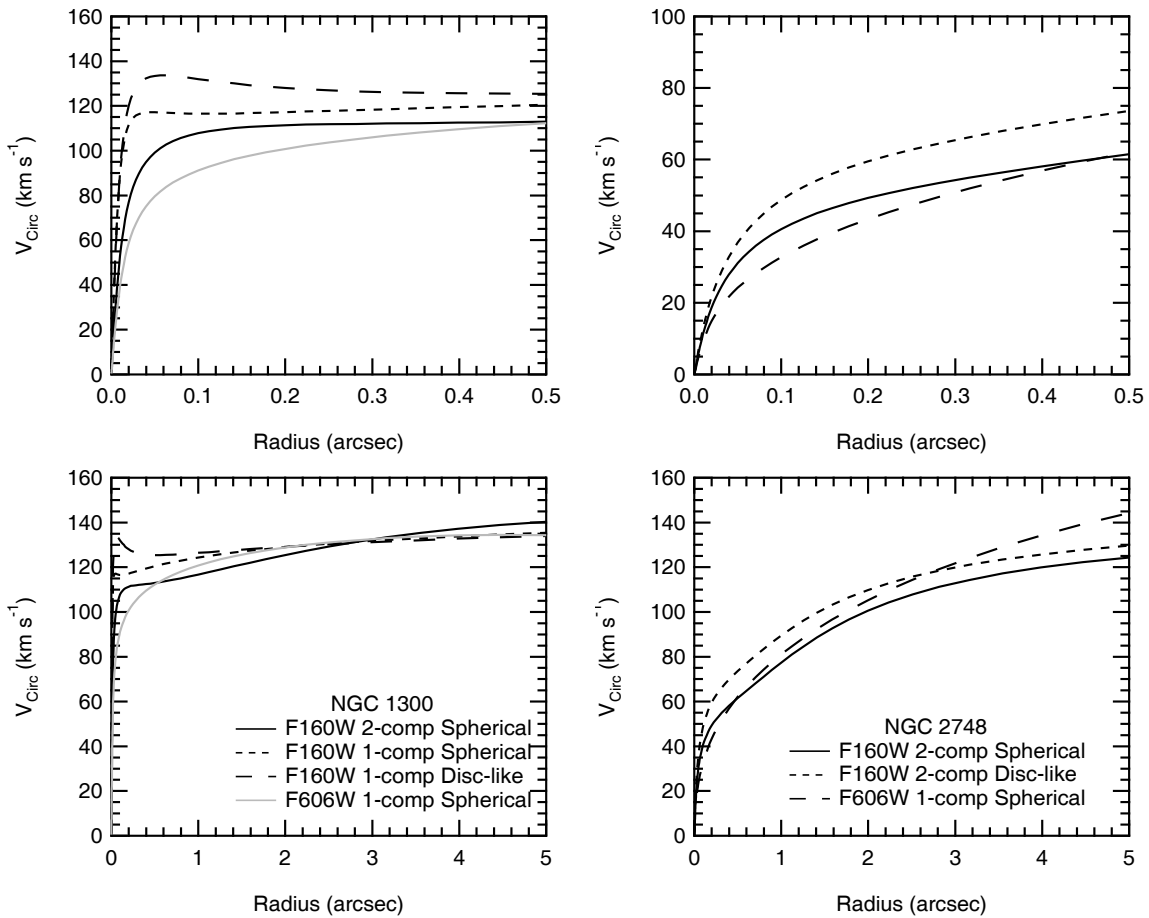

Figure 3. Circular velocities in the principal plane of the potential derived from the density models determined from the F160W and F606W images for NGC 1300 (left panels) and NGC 2748 (right panels). The lower panels show the same circular velocities as the top panels, but over a larger radial range. The velocities are calculated using the mass-to-light ratios derived in fitting the emission-line rotation curves (see later): these are as follows; NGC 1300: $\Upsilon_{F160W}$ (2-comp Sph) = 0.66, $\Upsilon_{F160W}$ (1-comp Sph) = 0.73, $\Upsilon_{F160W}$ (1-comp Disc) = 0.70 and $\Upsilon_{F606W}$ (2-comp Sph) = 2.3; NGC 2748: $\Upsilon_{F160W}$ (2-comp Sph) = 0.81, $\Upsilon_{F160W}$ (2-comp Disc) = 1.18 and $\Upsilon_{F606W}$ (1-comp Sph) = 6.8.

The F606W image of NGC 2748 exhibits significant extinction, southeast of the nucleus. The nucleus may be only partially visible; given the non-smooth central morphology, it is difficult to interpret exactly where the photometric centre lies. We performed significant masking when fitting ellipses to the isophotes; the resultant profile is not particularly smooth, probably caused by the prominent dust lanes, but is satisfactorily modelled using a single V98 component. The bottom right panel of Figure 2 shows there to be significant residuals in the model: however, we note that one will always obtain notable residuals when fitting a smooth model to a non-smooth profile. The model is satisfactory in matching the overall shape of the observed surface brightness profile.

The optimised parameters for each density model are listed in Table 2 with the associated value of the reduced $\chi^2$.

### 4.2 Kinematics

The reduced two-dimensional spectra were then analysed using the tasks LONGSLIT and DIPSO in the STARLINK FIGARO package. For each row we have fitted the emission lines with single Gaussians, approximating the continuum with a linear model. In the regions of extended emission, we have binned the spectra in LONGSLIT for two reasons: the signal-to-noise ratio of the data is vastly improved and we are not concerned with reproducing the small-scale 'wiggles' in the extended rotation curves that are probably due to spiral waves, local gas motions etc. Binning the extended rotation curves also has the advantage of dramatically reducing the computational time used in our models. In the central regions of emission, we have used DIPSO due to the added facility to constrain the parameters of the Gaussians during the fit. We have concentrated primarily on the Hα+[NII] feature due to its superior signal-to-noise ratio (compared with the [SII] doublet). Where possible we have fitted three Gaussians constrained to have the correct separations (accounting for their redshift) and the same full width at half maxima. In the case of the [NII] doublet we have also constrained the intensity ratio to be 3:1. A selection of typical resultant fits for NGC 1300 and NGC 2748 are displayed in Figure 4.

The extended regions of emission in NGC 1300 have been binned in bins of 10 rows (0.5″ in NUC, 1″ in OFF1/2). The extended rotation curve (Figure 5) is visible only on one side of the nucleus, with the exception of the OFF2 slit where binning the data has enabled us to measure one lone point on the opposite side to the rest of the data. The NUC and OFF1 central rotation curves show a steep gradient (peak to peak of ~ 200 km s$^{-1}$) compared with that of the extended rotation curve in OFF2. The OFF2 slit shows a strange central rotation curve in which the velocity has the same sense either side of the nucleus. This forms a 'smile-shaped' curve and we have been unable to model this feature, simultaneously with the additional rotation curves. We assume that this region is locally perturbed and is of little use as a tracer of mass. Therefore, we discard these velocity points when modelling the rotation curves.

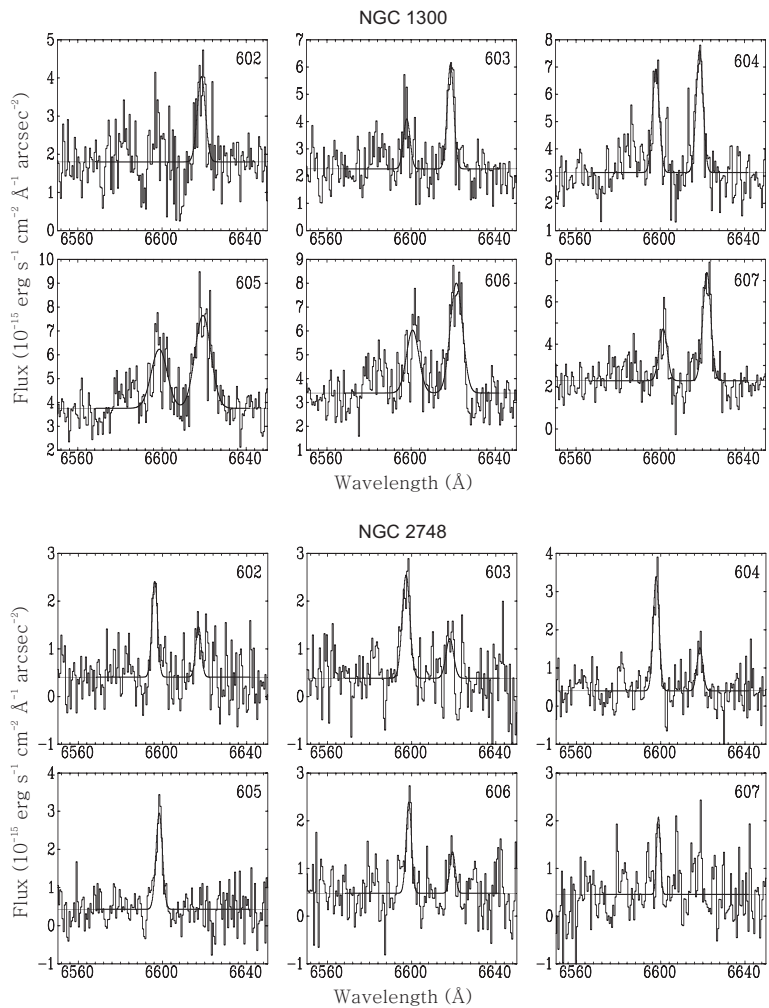

Figure 4. NGC 1300: A collection of typical Gaussian fits to emission lines in the nuclear regions (at radii between ~ ± 0.15″) of the NUC datasets of NGC1300 and NGC 2748. Where possible we fit constrained Gaussians, to both Hα (6565Å) and [NII] (6584Å). The smooth solid lines show the final fits to the observed spectrum.

The rotation curves derived for NGC 2748 are well sampled (Figure 5). Binning the regions of extended emission in 1″ bins enables us to trace the rotation curves over a spatial range of ~ 20″ (~ 2.25 kpc). The central regions exhibit strange rotation curves: over pixels ~ 590 to 610 (295 to 305 in the OFF1/2 rotation curves) there is a steep gradient of rotation (peak to peak ~ 150 km s$^{-1}$), perhaps indicative of a black hole. However, from pixels ~ 610 – 620, the velocities decline markedly. Again, we find that we cannot simultaneously reproduce these parts of the rotation curves in the three slits and thus, they are not used when modelling the rotation curves.

We adopt the notation that the binned sections of the rotation curves are termed the *extended rotation curves*, whilst the single row fits (i.e. in the nucleus) are termed the *central rotation curves*.

The ground-based rotation curves are plotted in

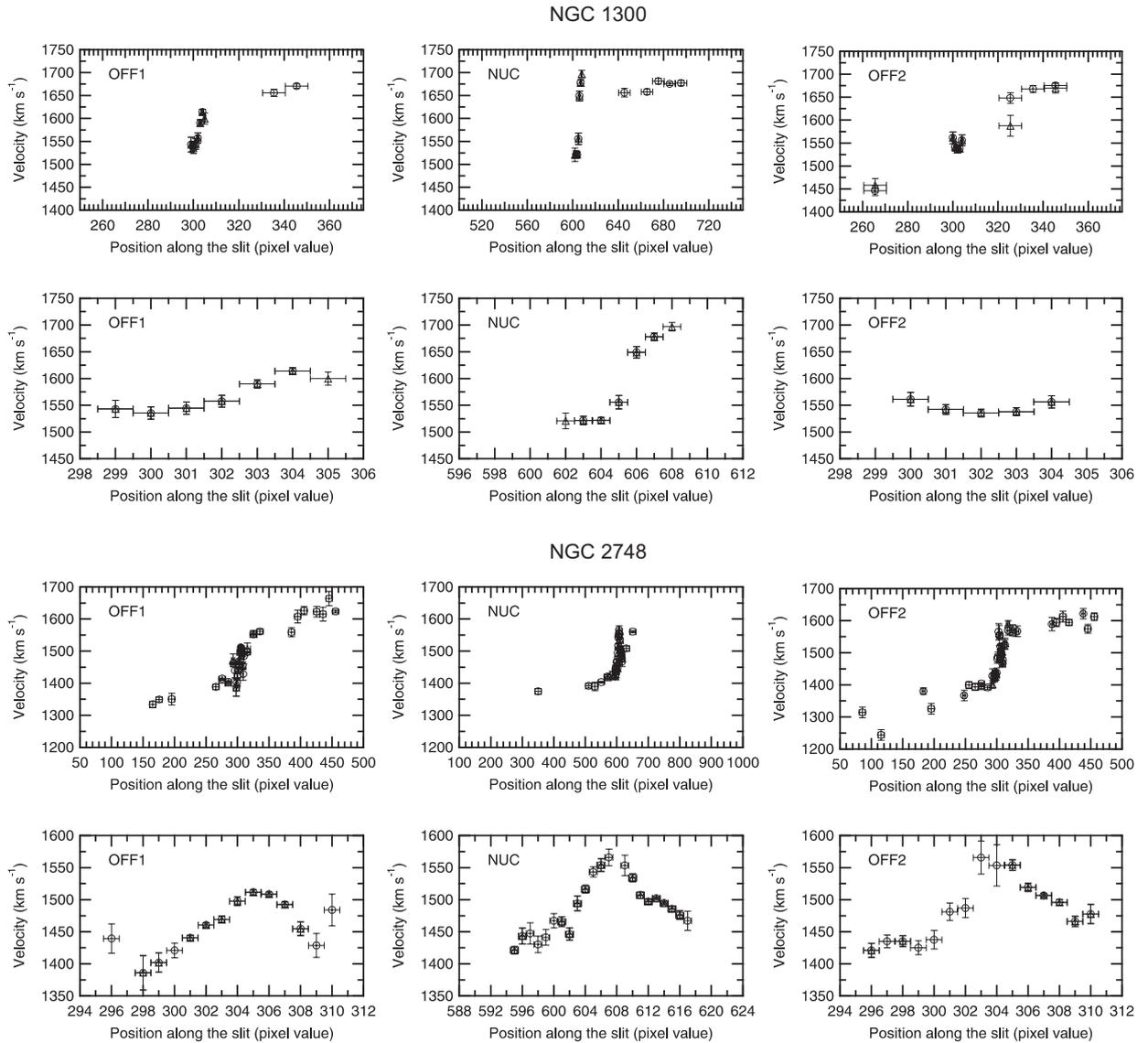

Figure 5 *HST/STIS* rotation curves. Expanded views of the central regions of the rotation curves are displayed below each full rotation curve. The OFF1/2 rotation curves are shown over the same radial range as the NUC rotation curves. Circles represent Hα (6563Å) and triangles represent [NII] (6584Å). The position of the brightest continuum flux (as determined in target acquisition) corresponds to NUC pixel rows 605 for both NGC 1300 and NGC 2748, while the position of the brightest emission-line flux corresponds to pixel row 605 in NGC 1300 and pixel row 613 for NGC 2748.

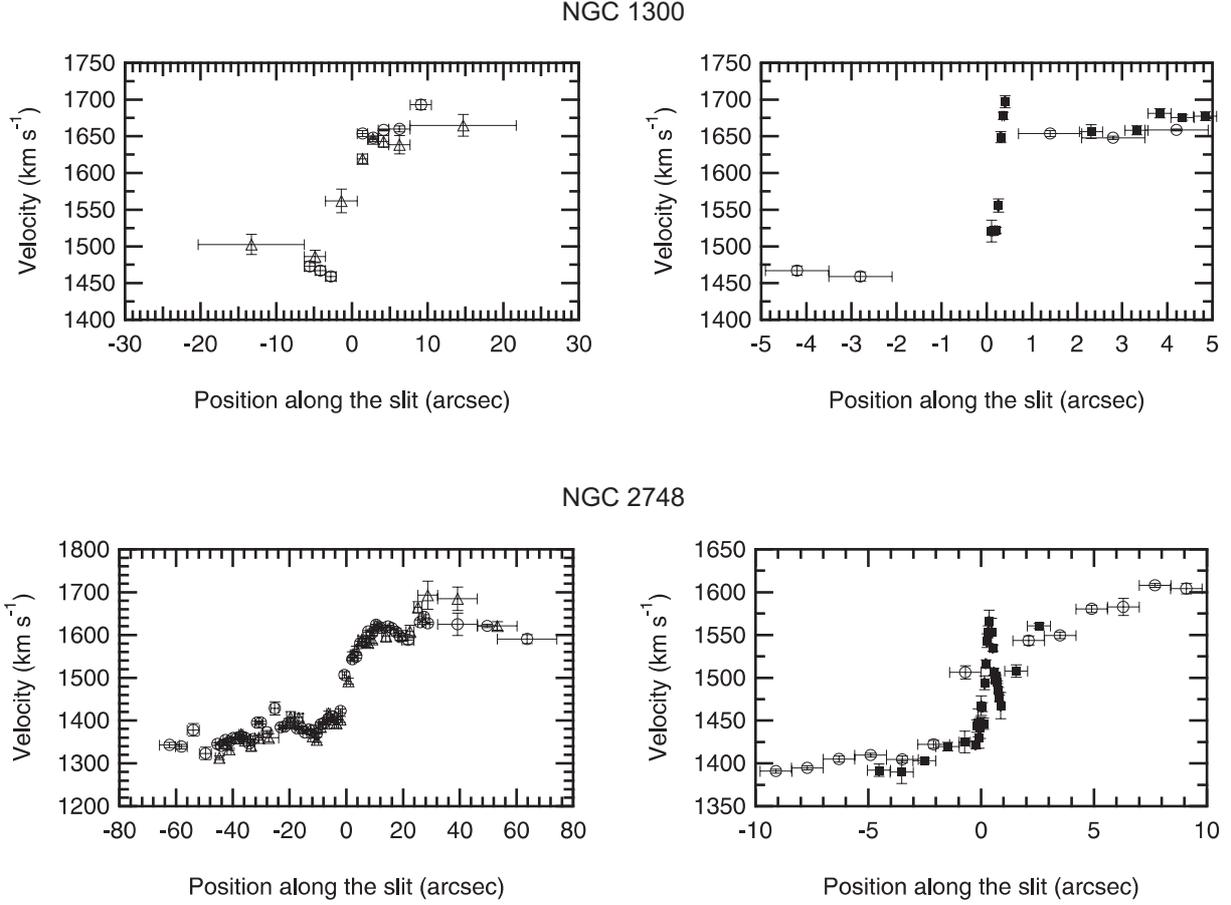

Figure 6. Ground-based rotation curves (left). Expanded views of the central regions of the ground-based Hα rotation curves are overlaid on the *STIS* NUC slit, weighted mean velocity rotation curves (right). Circles represent Hα (6563Å), triangles represent [NII] (6584Å) and solid squares represent the *STIS* rotation curves.

Figure 6 where we also overlay the corresponding *STIS* rotation curves (derived from the weighted mean of all emission lines) on the ground-based Hα rotation curve. The ground-based rotation curves for both NGC 1300 and NGC 2748 were obtained along the major axis of each galaxy. When overlaid, the *STIS* NUC rotation curves agree well with the ground based rotation curve: Figure 6 shows how the steep central gradient of rotation is only observed with *HST's* superior resolution.

### 4.3 Modelling the emission-line flux distribution

We use the reference pixels taken from the CRPIX2 values in the FITS header of each spectrum to ensure that the spectra are correctly aligned. We also take into account the fact that the spectra are shifted and then realigned in order to enable accurate cosmic ray rejection.

The emission-line flux distributions along each slit are used to estimate a two-dimensional image (or map) of the emission-line flux distribution. Modelling the emission-line flux distribution is only important in regions of strong velocity gradients. As such, we neglect the effect of the flux distribution of the extended rotation curves and only model the flux distributions in the central rotation curves of all three slits simultaneously. We model the flux from the most completely sampled emission-line or, in cases of equivalent sampling, we adopt Hα as the [NII] doublet (which is collisionally excited) might be more affected by shocks.

The optimised parameters for each analytical component for each emission-line flux model are listed in Table 3. The resultant models are overlaid on the observed emission-line flux distributions in Figure 7.

*NGC 1300*

The NUC emission-line flux profile of NGC

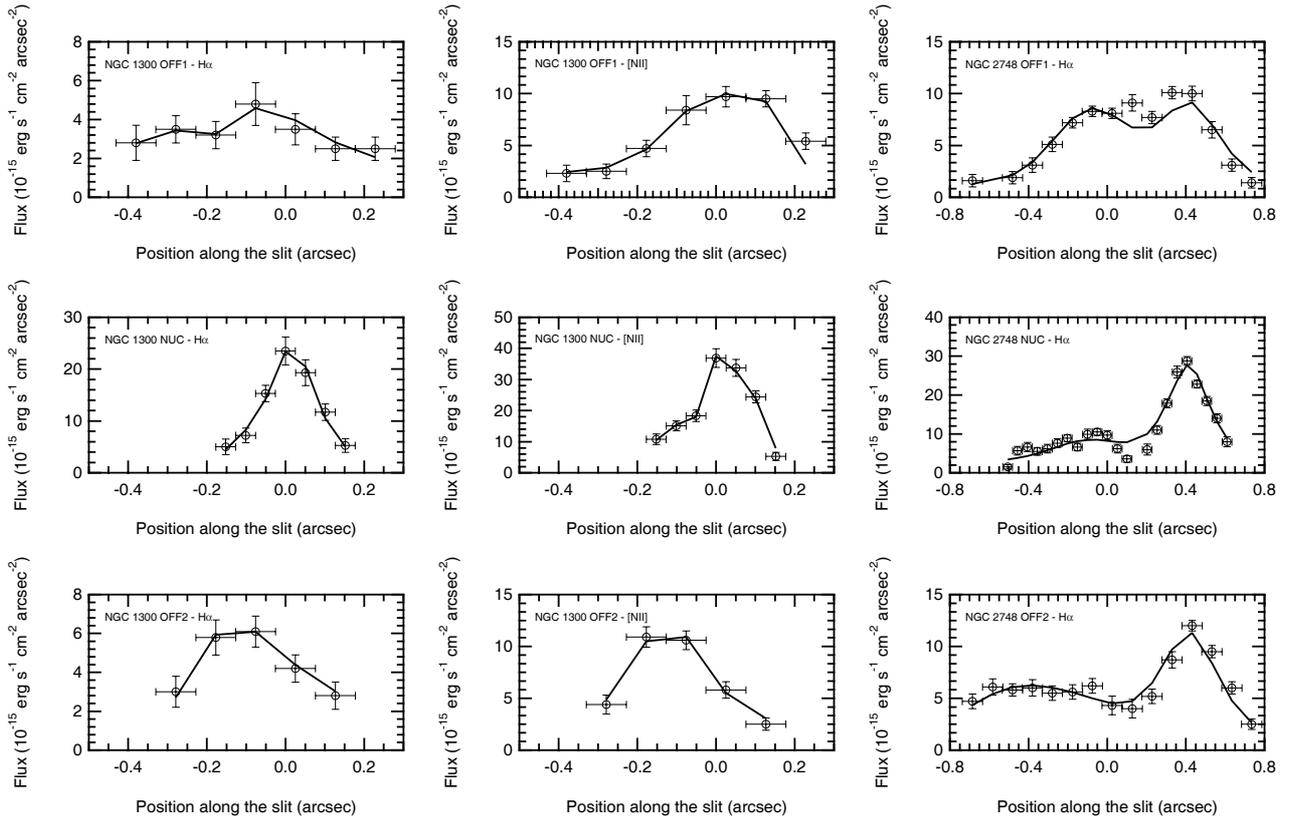

Figure 7 Model flux distributions (solid lines) overlaid on the observed data.

1300 shows a strong peak in the central region. The flux distribution in the OFF2 slit is similarly peaked but much weaker; whilst in the OFF1 slit, the emission has a less prominent central peak. The given slit centres mean that the peak of the flux distribution in each of the slits is not aligned and thus a single peaked model will not be sufficient to reproduce the observed profile. Instead a combination of 4 exponentials and a constant baseline is found to be the best workable model for reproducing the observed flux distribution. The [NII] emission-line profile is significantly different to the Hα emission-line profile in shape. Consequently, we also modelled the [NII] flux profile which is used to investigate how a different flux map affects the derived black hole mass.

*NGC 2748*

In NGC 2748, the emission-line flux distribution peaks on the NUC and OFF2 slit at around at radius of ~ 0.4″. The peak of flux in the OFF1 slit is much broader and is centred on a radius of ~ 0.2″. We adopt a synthetic model of an exponential, two Gaussians and a constant baseline to reproduce the observed distribution.

### 4.4 Modelling the rotation curves

We model the rotation curves using two strategies following Marconi et al. (2003). We begin using the *standard approach* in which we assume that the full gas disc is coplanar (the inclination is fixed at its large-scale value) and that the mass-to-light ratio of the stellar component is also constant with radius. Theoretically, the use of three parallel slits should enable us to constrain the angle between the slit and line of nodes. However, this relies on asymmetries in the central regions of the rotation curves and small changes in the shape of the rotation curve. Such signatures can be difficult to pin down (although not impossible) given the intrinsic small-scale variations in the rotation curves. Consequently we will run models where the angle between the slit and line of nodes is both fixed to its large-scale value and left as a free parameter of the fit.

Our second approach, the *alternative approach*, relaxes these constraints: we fit only the central rotation curves allowing the mass-to-light ratio or the inclination to vary. We note that these parameters are still assumed to be constant over the region to which we fit. We fit the extended rotation curves (with fixed geometry and inclination based on large-scale values) to derive the mass-to-light ratio of the system, and then fit the central rotation curves, keeping the mass-to-light ratio fixed at the value derived from fitting the extended rotation curves, but allowing the inclination to vary. Conversely, we keep the inclination fixed at its large-scale value, but allow for a different mass-to-light ratio. In all cases the geometry of the disc is unconstrained.

We use the extended rotation curves to estimate any offsets in the systemic velocity (due to uncertainties in wavelength calibration) of the OFF1/2 spectra with respect to the NUC spectrum. Modelling of NGC 1300 suggests that such offsets are approximately -7 and -4 km s$^{-1}$ for the OFF1 and OFF2 slits, respectively: for NGC 2748 we find them to be approximately -1 and -4 to -5 km s$^{-1}$ (depending on which bulge model is used). We note that in NGC 1300 we observe only one velocity point on the approaching side of the disc, in the OFF2 rotation curve. Consequently, the velocity offsets for NGC 1300 are not particularly robust; however comparing the OFF2 extended rotation curve with the ground-based rotation curve we see that the single data point is not spurious.

*NGC 1300*

To satisfactorily model both the central and extended rotation curves simultaneously, a shift in systemic velocity between them in the range 28 to 35 km s$^{-1}$ is required. We attribute this shift to the fact the systemic velocity is constrained by only the single data point on the approaching side of the disc. The uncertainty of the Gaussian fits to this particular point is of order 8 km s$^{-1}$ which, combined with the degree of modulation seen in the extended rotation curves of other galaxies in our sample, satisfactorily accounts for this shift.

Table 4 lists the optimised parameters derived in a number of models fitted to the rotation curves of NGC 1300. We define the best base model to be that which uses the spherical density model comprising the two components necessary to reproduce the complete surface brightness profile derived from the F160W image, and the H$\alpha$ flux map. The complete list of models investigated is:

1. a two-component spherical density model (based on the F160W photometry) and the flux map derived from the H$\alpha$ flux profiles;
2. a two-component spherical density model (based on the F160W photometry) and the flux map derived from the [NII] flux profiles;
3. a one-component spherical density model (based on the F160W photometry) and the flux map derived from the H$\alpha$ flux map;
4. a one-component 'disc-like' ($q = 0.1$) density model (based on the F160W photometry) and the flux map derived from the H$\alpha$ flux map;

Table 3. Optimised parameters for individual analytic components of the model emission-line flux maps.

| Galaxy | Emission line | Component | Offset cross-dispersion (arcsec) | Offset dispersion (arcsec) | Relative intensity | Scale radius (arcsec) |
|---|---|---|---|---|---|---|
| NGC 1300 | H$\alpha$ | Exponential | 0.02 | 0.00 | 155.9 | 0.03 |
| | | Exponential | -0.14 | -0.19 | 49.0 | 0.03 |
| | | Exponential | -0.31 | 0.21 | 11.1 | 0.04 |
| | | Exponential | -0.10 | 0.18 | 27.3 | 0.02 |
| | | Constant | N/A | N/A | 1.7 | N/A |
| NGC 1300 | [NII] | Exponential | 0.01 | 0.02 | 2693.0 | 0.01 |
| | | Gaussian | -0.13 | -0.12 | 30.0 | 0.15 |
| | | Gaussian | 0.08 | 0.09 | 1631.2 | 0.02 |
| | | Gaussian | -0.11 | 0.16 | 1355.8 | 0.01 |
| | | Constant | N/A | N/A | 2.3 | N/A |
| NGC 2748 | H$\alpha$ | Exponential | 0.42 | -0.01 | 61.2 | 0.1 |
| | | Gaussian | -0.08 | 0.13 | 9.6 | 0.4 |
| | | Gaussian | -0.45 | -0.42 | 7.7 | 0.6 |
| | | Constant | N/A | N/A | 0.8 | N/A |

5. a one-component spherical density model (based on the F606W photometry) and the flux map derived from the Hα flux map.

Figure 8 shows that a model without a black hole clearly fails to reproduce the steep central gradient of rotation observed in our *SITS* rotation curves of NGC 1300. This view is supported by the fact that the $\chi^2$ is much larger when compared with fits to models including a black hole (see Table 4). Using the best base model we find a black hole of mass of ~ $7.4 \times 10^7$ $M_\odot$ is required to reproduce the observed rotation curves. It is also comforting that the optimised black hole mass is almost identical when the disc is fixed coplanar and when the angle between the slit and the line of nodes is optimised to 1° (an offset of only 3° to the large-scale line of nodes). The

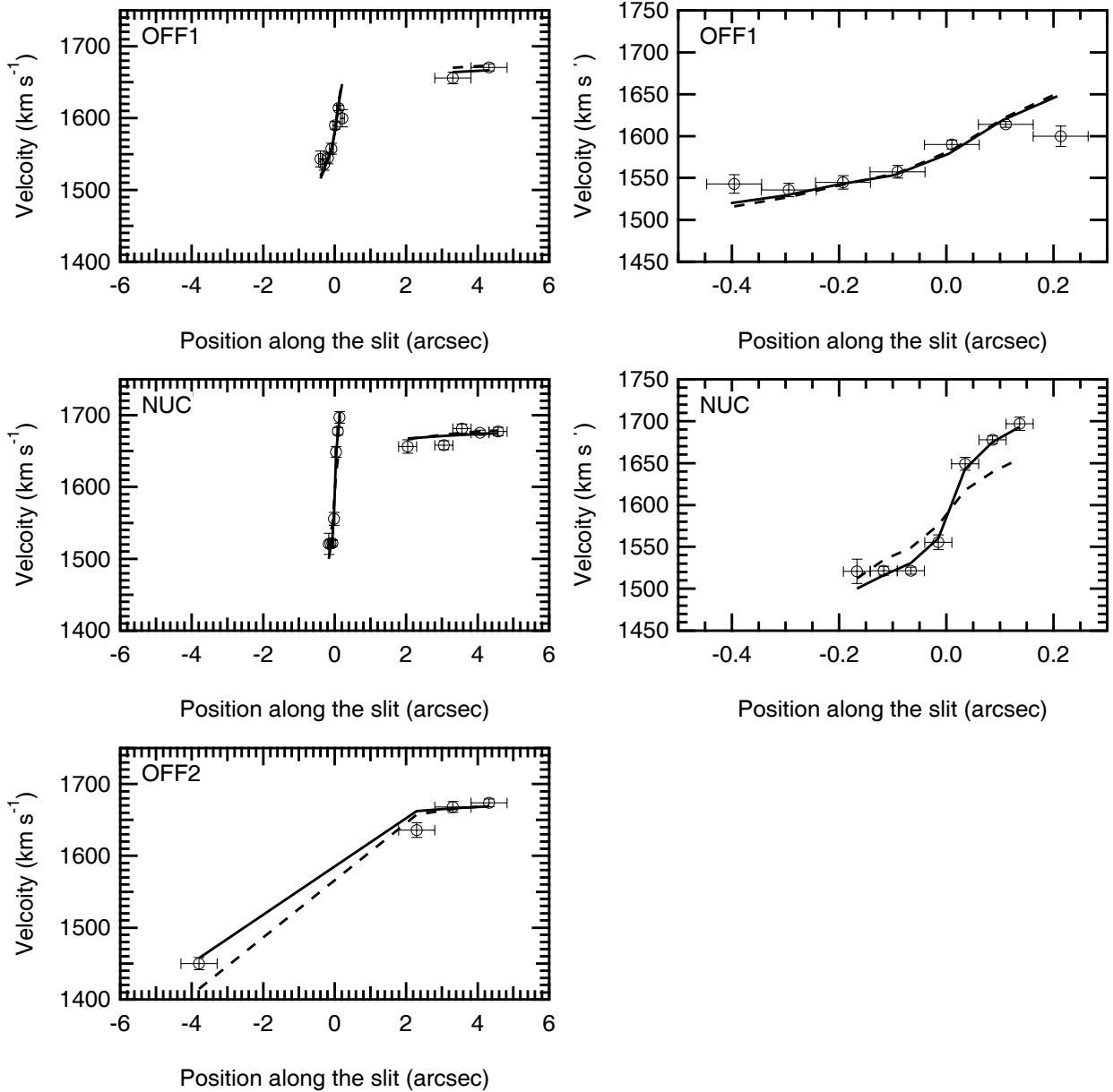

Figure 8. Model rotation curves derived for models with no black hole (dashed line) and a black hole of mass ~ $7 \times 10^7$ $M_\odot$ (solid line) overlaid on the observed rotation curves of NGC 1300. Expanded views of the central regions are displayed on the right. Both models use the mass density profile based on the spherical two-component fit to the F160W surface brightness profile and the flux map derived from the Hα emission-line profiles. The line that connects the points is used purely as a visual aid. The central and extended model rotation curves are not connected by the line to indicate that they are kinematically distinct. All models were based on the assumptions of the *standard approach* as defined in section 4.4.

*Atkinson et al.*

impact parameter and offset in the along the slit direction are both small (less than a pixel) suggesting that the spacecraft was indeed correctly centred on the kinematic centre of NGC 1300. The other base models return similar geometric parameters, and black hole mass estimates compared with the best base model. The range of black hole masses derived from all of the models range from ~ $4.3 \times 10^7$ to $7.9 \times 10^7$ $M_\odot$, however, the optimised model based on the best base model has a significantly lower value for $\chi^2$.

Estimating the uncertainty in our best estimate of the black hole mass is accomplished by creating a grid of black hole mass and mass to light ratio values (around the optimised values of the best base model) and optimising a model at each grid point – the result being $\chi^2$ ellipses (displayed in Figure 10). The high degree of modulation in the rotation curves means that any simple disc model will fail to produce reduced $\chi^2$ values close to unity. As such we follow the approach given by Barth et al. (2001) and also used in our analysis of NGC 4041, and have quadratically added a constant (10.8 km s$^{-1}$) to the uncertainties in velocity at each point along the slit. This constant is chosen to ensure that the best fitted model gives a reduced $\chi^2 \sim 1$. In this fashion, we are rescaling the $\chi^2$ ellipses. The resultant $\chi^2$ contours are widened by this method and show the statistical uncertainties (at the 95 % confidence level) in the black hole mass to be $6.6^{+6.3}_{-3.2} \times 10^7\ M_\odot$. We note that this value of black hole mass differs slightly from that derived originally (7.4 × 10$^7\ M_\odot$). This discrepancy arises from the fact that by adding a constant to the velocity uncertainties, and thus widening the confidence intervals of the $\chi^2$ ellipses, we have made the minimum in parameter space shallower. There are, in fact two grid points which have the lowest value of $\chi^2$. We quote this latter black hole mass as the best estimate (although there is little difference between them). A glance at Table 4 shows that the optimised black hole masses based on the other base models are within the range of uncertainty derived above.

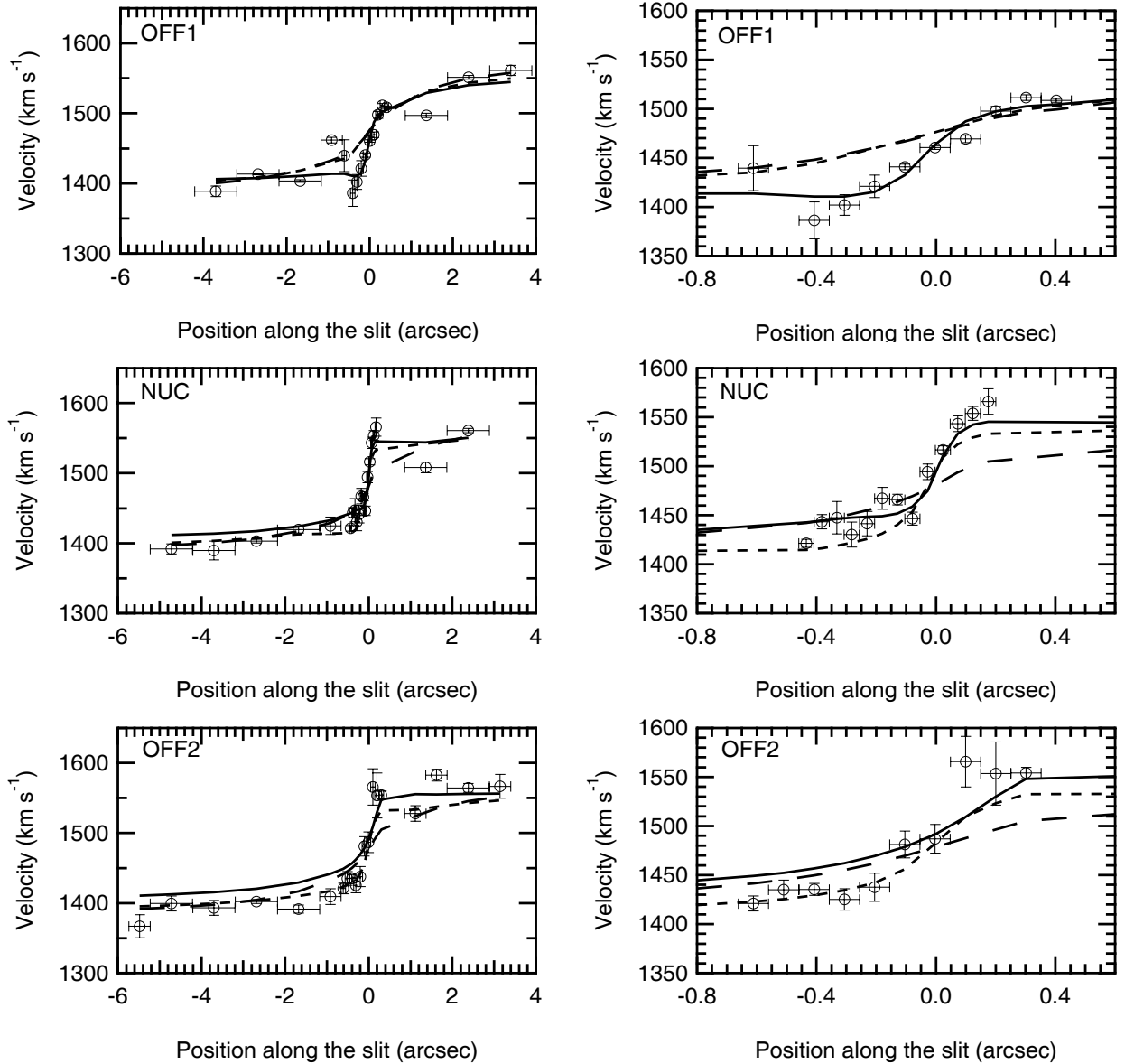

Figure 9. Model rotation curves derived for models with no black hole (long-dashed line), a black hole of mass ~ 5 × 10$^7$ $M_\odot$ with a coplanar disc (short-dashed line) and a black hole of mass ~ 4 × 10$^7$ $M_\odot$ with the angle between the slit and the line of nodes optimised at -23° (solid line), overlaid on the observed rotation curves of NGC 2748. Expanded views of the central regions are displayed on the right. All models use the mass density profile based on the spherical two-component fit to the F160W surface brightness profile and the flux map derived from the Hα emission-line profiles. The line that connects the points is used purely as a visual aid. All models were based on the assumptions of the *standard approach* as defined in section 4.4.

Table 4. Optimised parameters for velocity field models used in fitting the rotation curves using the *standard approach* defined in section 4.4. Photometry band indicates the image used to derive the surface brightness profile, which was then used to infer the density model (the number of component in the density model are also listed); $q$ is the axial ratio of the density model; Flux map indicates the emission-line profile modelled to infer the flux map; Log $M_{BH}$ is the log of black hole mass; $\Upsilon$ is the mass-to-light ratio of the stellar component; $S_0$ is the slit centre; $b$ is the impact parameter; $\theta$ is the angle between the slit and the line of nodes; $i$ is the inclination; $V_{sys}$ is the systemic velocity of the central rotation curves; $\Delta V_{sys}$ is the shift in systemic velocity of the extended rotation curves (with respect to the central rotation curves), and $\chi^2$ and $\chi^2_{red}$ are the optimised values (the subscript 'red' indicating the reduced $\chi^2$). A † indicates that the parameter was fixed to that value during the fit.

| Galaxy | Photometry band | $q$ | Flux map | Log $M_{BH}$ ($M_\odot$) | $\Upsilon$ ($\Upsilon_\odot$) | $S_0$ (arcsec) | $b$ (arcsec) | $\theta$ (°) | $i$ (°) | $V_{sys}$ (km s$^{-1}$) | $\Delta V_{sys}$ (km s$^{-1}$) | $\chi^2$ | $\chi^2_{red}$ |
|---|---|---|---|---|---|---|---|---|---|---|---|---|---|
| NGC 1300 | F160W: 2-comp | 1† | Hα | 0.00† | 1.19 | -0.015 | -0.014 | 17 | 49† | 1583 | -35 | 232.18 | 11.6 |
|  | F160W: 2-comp | 1† | Hα | 7.88 | 0.66 | -0.022 | -0.036 | 4† | 49† | 1598 | -31 | 54.44 | 2.7 |
|  | F160W: 2-comp | 1† | Hα | 7.87 | 0.65 | -0.022 | -0.033 | 1 | 49† | 1602 | -35 | 53.99 | 2.8 |
|  | F160W: 2-comp | 1† | [NII] | 7.63 | 0.65 | -0.012 | 0.013 | 4† | 49† | 1598 | -30 | 92.23 | 4.6 |
|  | F160W: 2-comp | 1† | [NII] | 7.78 | 0.62 | -0.012 | 0.009 | -3 | 49† | 1606 | -36 | 89.52 | 4.7 |
|  | F160W: 1-comp | 1† | Hα | 7.86 | 0.73 | -0.022 | -0.036 | 4† | 49† | 1598 | -29 | 63.73 | 3.2 |
|  | F160W: 1-comp | 1† | Hα | 7.84 | 0.72 | -0.021 | -0.033 | 3 | 49† | 1599 | -29 | 63.75 | 3.4 |
|  | F160W: 1-comp | 0.1† | Hα | 7.82 | 0.70 | -0.022 | -0.038 | 4† | 49† | 1598 | -29 | 67.38 | 3.4 |
|  | F160W: 1-comp | 0.1† | Hα | 7.84 | 0.72 | -0.023 | -0.039 | 0 | 49† | 1605 | -38 | 66.83 | 3.5 |
|  | F606W: 1-comp | 1† | Hα | 7.90 | 2.31 | -0.021 | 0.034 | 4† | 49† | 1597 | -30 | 61.06 | 3.1 |
|  | F606W: 1-comp | 1† | Hα | 7.90 | 2.27 | -0.021 | -0.033 | 3 | 49† | 1599 | -37 | 60.89 | 3.2 |
| NGC 2748 | F160W: 2-comp | 1† | Hα | 0.00† | 0.40 | 0.053 | 0.032 | 0† | 73† | 1481 | - | 994.29 | 18.8 |
|  | F160W: 2-comp | 1† | Hα | 7.73 | 0.32 | 0.096 | -0.060 | 0† | 73† | 1478 | - | 746.57 | 14.4 |
|  | F160W: 2-comp | 1† | Hα | 7.64 | 0.81 | 0.073 | 0.064 | -23 | 73† | 1481 | - | 714.00 | 14.0 |
|  | F160W: 2-comp | 0.1† | Hα | 7.71 | 0.41 | 0.098 | -0.060 | 0† | 73† | 1478 | - | 752.39 | 14.5 |
|  | F160W: 2-comp | 0.1† | Hα | 7.53 | 1.18 | 0.080 | 0.070 | -25 | 73† | 1480 | - | 726.01 | 142 |
|  | F606W: 1-comp | 1† | Hα | 7.79 | 2.8 | 0.097 | -0.064 | 0† | 73† | 1478 | - | 744.58 | 14.3 |
|  | F606W: 1-comp | 1† | Hα | 7.72 | 6.8 | 0.076 | 0.064 | -23 | 73† | 1482 | - | 687.17 | 13.5 |

The part of Table 5 labelled *NGC 1300 Extended* lists the optimised parameters of the models fitted to the extended rotation curves. These model fits are primarily used to derive the mass-to-light ratio of the system when modelling the rotation curves using the *alternative approach*. The part of Table 5 labelled *NGC 1300 Central* lists the optimised parameters of the various models fitted to only the central rotation curves. We can clearly see that models (without a black hole) which allow the inclination or mass-to-light ratio to differ from their respective large-scale values result in a value of $\chi^2$ much larger than the model which fixes these parameters and allows the inclusion of a black hole. Note that for conciseness, we have listed only one model which includes a black hole, using best base model. In all cases, the models which do not include a black hole result in reduced $\chi^2$ values greater than 10, compared with ~ 4 for that model including the black hole. In the left panels of Figure 11 we show the resultant model rotation curves for the first three models; the best fitting model is clearly that in which the inclination and mass-to-light ratio are kept constant while the mass of a black hole is optimised, increasing the robustness of our original black hole mass estimate for NGC 1300.

*NGC 2748*

For NGC 2748, we define the best base model as that which uses the spherical two-component density model based on the F160W surface brightness profile. The complete list of base models used in fitting NGC 2748 is:

1. a spherical two-component density model derived from the F160W image;
2. a two-component density model with the extended component being 'disc-like', derived from the F160W image;
3. a one-component spherical density model derived from the F606W surface brightness profile.

All models are based on the Hα flux map (there is little difference in shape between the Hα and [NII] flux maps). The optimised parameters are listed in Table 4 for NGC 2748.

In order to reproduce the observed central rotation curves of NGC 2748, a black hole of mass $4.4 \times 10^7$ $M_\odot$ is required by the best base model. Models without a black hole yield unsatisfactory fits and return a larger $\chi^2$ than those with a black hole included (see Figure 9 and Table 4). A good fit to the central rotation curves requires the angle between the slit and line of nodes to be ~ -23°, however, if the disc is coplanar, we expect the slit to lie along the kinematic line of nodes (i.e. $\theta = 0°$). Performing the same error analysis used in the case of NGC 1300 (the constant added quadratically to the velocity uncertainties was 19 km s$^{-1}$) we find the statistical uncertainties (at the 95% confidence level) in black hole mass yield a mass of $4.4^{+3.5}_{-3.6} \times 10^7$ $M_\odot$. In deriving the $\chi^2$ ellipses for NGC 2748, we note that the model which yields the lowest value of $\chi^2$ is that for which the mass-to-light ratio of the stellar component differs from the original optimised value. The difference is reasonably small: the original mass-to-light ratio was ~ 0.8 whilst the mass-to-light ratio yielding the lowest rescaled $\chi^2$ is ~ 1.1. This does not affect our black hole mass estimate: we simply mention it for clarity.

Inspection of Table 4 shows that the optimised black hole masses based on the other base models all lie within the uncertainties of the optimised black hole mass derived using the best base model. The reduced $\chi^2$ values are reasonably large (> 10) for all of the models used in fitting the rotation curves of NGC 2748. However, these large values are caused by the high degree of modulation which the rotation curves exhibit, suggesting that a reduced $\chi^2$ larger than one is inevitable. Consequently, we feel that the large values do not invalidate our black hole mass estimates: indeed we have accounted for the large values of reduced $\chi^2$ by rescaling them when calculating the uncertainties in our black hole mass estimates.

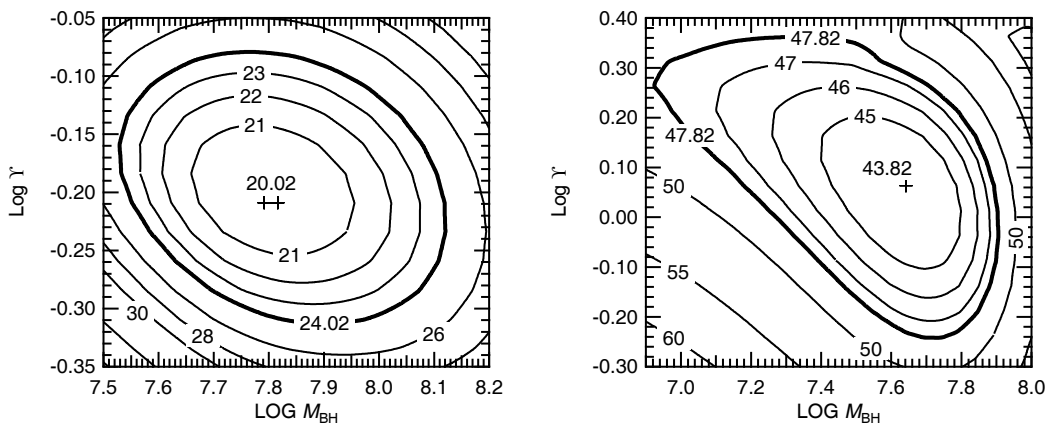

Figure 10 Contours of equal values of $\chi^2$ determined by a grid of black hole mass and mass-to-light ratio for NGC 1300 (left) and NGC 2748 (right). The cross indicates the value of $\chi^2$ for the best fitting models. For NGC1300 this model has 21 degrees of freedom, whilst the best fitting model for NGC 2748 has 53 degrees of freedom. The thick line contour represents the 95% confidence level in the determined parameters.

The optimised parameters for the various models based on the *alternative approach* are listed in Table 5. Due to the high inclination of NGC 2748 (~ 73°), varying the inclination in modelling the rotation curves is unlikely to increase the amplitude of the model rotation curves. Indeed, this is the case; Table 5 shows that in all cases when optimised, the inclination is lower than that of the large-scale disc and Figure 11 shows the model clearly fails to reproduce the observed rotation curves.

Unlike the models of NGC 1300, we see that allowing the mass-to-light ratio of the stellar component to vary we can satisfactorily reproduce the observed central rotation curves without the need for a black hole ($\chi^2 \sim 102$). However, this is compared to the model with a black hole, in which the inclination is fixed to its large-scale value and the mass-to-light ratio of the bulge is held fixed at the value derived in fitting to only the extended rotation curves ($\chi^2 \sim 171$). If we allow both the black hole mass and the mass-to-light ratio of the bulge to vary during the fit, we find that a black hole of mass $\sim 2 \times 10^7\ M_\odot$ yields the lowest $\chi^2$ (~96). The important point to notice here is that, for both models (from the best base model) the optimised mass-to-light ratios are rather large compared to those that are typically found. Bell & de Jong (2001) showed the stellar mass-to-light ratio is correlated with galaxy colour. For NGC 2748, we adopt an effective B-V colour of 0.82 (LEDA). We then find that, for the wide range of models described by Bell & de Jong (e.g. closed box, infall, outflow, etc.), the predicted *H* band mass-to-light ratio lies in the range 0.82 to 0.95. Moriondo, Giovanardi, & Hunt (1998) derive typical mass-to-light ratios for early-type spirals ranging from roughly 0.1 to 0.9 in the *K* band and ~ 0.1 to 1.9 in the *J* band. Maraston (1998) suggests, from evolutionary synthesis of stellar populations, that even for a Salpeter IMF population of age 15 Gyr, the largest *K* band mass-to-light ratio is only 1.3.

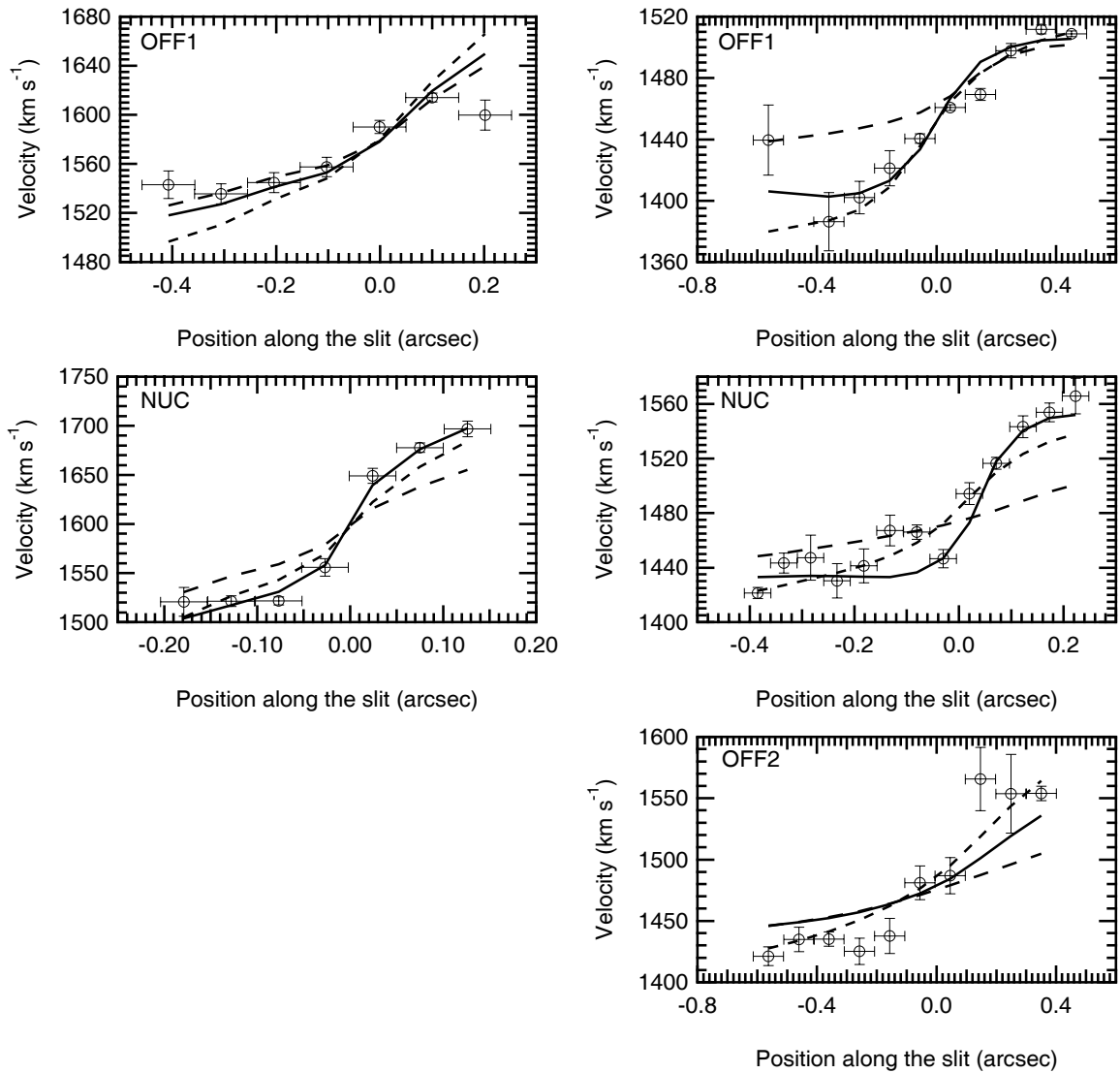

Figure 11. Central rotation curves of NGC 1300 (left) and NGC 2748 (right) derived for models based on the *alternative approach* (defined in section 4.4). The solid lines indicate rotation curves for models where the mass-to-light ratio and inclination were held fixed at those values for the large-scale data and a black hole whose mass is a free parameter of the fit, is included. The short dashed lines represent rotation curves derived for models with no black hole, where the mass-to-light of the stellar component is a free parameter of the fit. The long dashed lines represent rotation curves of models with no black hole, where the inclination of the disc is a free parameter of the fit.

Table 5. Optimised parameters for velocity field models used in fitting the rotation curves using the *alternative approach* defined in section 4.4. Photometry band indicates the image used to derive the surface brightness profile, which was then used to infer the density model (the number of component in the density model are also listed); $q$ is the axial ratio of the density model; Flux map indicates the emission-line profile modelled to infer the flux map; Log $M_{BH}$ is the log of black hole mass; $\Upsilon$ is the mass-to-light ratio of the stellar component; $S_0$ is the slit centre; $b$ is the impact parameter; $\theta$ is the angle between the slit and the line of nodes; $i$ is the inclination; $V_{sys}$ is the systemic velocity; and $\chi^2$ and $\chi^2_{red}$ are the optimised values (the subscript 'red' indicating the reduced $\chi^2$). A † indicates that the parameter were fixed to that value during the fit.

| Galaxy | | Photometry band | $q$ | Flux map | Log $M_{BH}$ ($M_\odot$) | $\Upsilon$ ($\Upsilon_\odot$) | $S_0$ (arcsec) | $b$ (arcsec) | $\theta$ (°) | $i$ (°) | $V_{sys}$ (km s$^{-1}$) | $\chi^2$ | $\chi^2_{red}$ |
|---|---|---|---|---|---|---|---|---|---|---|---|---|---|
| NGC 1300 | Extended | F160W: 2-comp | 1† | N/A | 0.00† | 0.75 | 0.00† | 0.00† | 4† | 49† | 1563 | 14.36 | 2.1 |
| | | F160W: 1-comp | 1† | N/A | 0.00† | 0.88 | 0.00† | 0.00† | 4† | 49† | 1562 | 19.37 | 2.8 |
| | | F160W: 1-comp | 0.1† | N/A | 0.00† | 0.84 | 0.00† | 0.00† | 4† | 49† | 1563 | 20.40 | 2.9 |
| | | F606W: 1-comp | 1† | N/A | 0.00† | 2.69 | 0.00† | 0.00† | 4† | 49† | 1563 | 19.86 | 2.8 |
| | Central | F160W: 2-comp | 1† | Hα | 0.00† | 1.87 | -0.027 | -0.025 | 2 | 49† | 1603 | 140.99 | 15.7 |
| | | F160W: 2-comp | 1† | Hα | 0.00† | 0.75† | -0.021 | -0.002 | 4 | 59 | 1598 | 243.78 | 27.1 |
| | | F160W: 2-comp | 1† | Hα | 7.91 | 0.75† | -0.023 | -0.040 | -4 | 49† | 1609 | 39.34 | 4.4 |
| | | F160W: 1-comp | 1† | Hα | 0.00† | 1.74 | -0.024 | -0.023 | 3 | 49† | 1605 | 127.20 | 14.1 |
| | | F160W: 1-comp | 1† | Hα | 0.00† | 0.88† | -0.022 | -0.003 | 3 | 59 | 1598 | 185.36 | 20.6 |
| | | F160W: 1-comp | 0.1† | Hα | 0.00† | 1.41 | -0.024 | -0.023 | 2 | 49† | 1604 | 108.97 | 12.1 |
| | | F160W: 1-comp | 0.1† | Hα | 0.00† | 0.84† | -0.021 | -0.004 | 4 | 60 | 1599 | 137.25 | 15.3 |
| | | F606W: 1-comp | 1† | Hα | 0.00† | 7.73 | -0.027 | -0.027 | 4 | 49† | 1605 | 188.55 | 21.0 |
| | | F606W: 1-comp | 1† | Hα | 0.00† | 2.69† | -0.021 | -0.001 | 5 | 60 | 1597 | 319.31 | 35.5 |
| NGC 2748 | Extended | F160W: 2-comp | 1† | N/A | 0.00† | 0.38 | 0.00† | 0.00† | 0† | 73† | 1480 | 377.33 | 18.9 |
| | | F160W: 2-comp | 0.1† | N/A | 0.00† | 0.48 | 0.00† | 0.00† | 0† | 73† | 1480 | 389.72 | 19.5 |
| | | F606W: 1-comp | 1† | N/A | 0.00† | 3.30 | 0.00† | 0.00† | 0† | 73† | 1480 | 381.85 | 19.1 |
| | Central | F160W: 2-comp | 1† | Hα | 0.00† | 2.68 | 0.122 | 0.077 | -27 | 73† | 1470 | 101.79 | 3.7 |
| | | F160W: 2-comp | 1† | Hα | 0.00† | 0.38† | 0.017 | 0.123 | -5 | 62 | 1482 | 522.18 | 18.7 |
| | | F160W: 2-comp | 1† | Hα | 7.87 | 0.38† | 0.084 | 0.063 | -16 | 73† | 1478 | 170.75 | 6.1 |
| | | F160W: 2-comp | 1† | Hα | 7.22 | 2.10 | 0.107 | 0.063 | -24 | 73† | 1473 | 96.07 | 3.6 |
| | | F160W: 2-comp | 0.1† | Hα | 0.00† | 2.86 | 0.121 | 0.075 | -26 | 73† | 1469 | 101.73 | 3.6 |
| | | F160W: 2-comp | 0.1† | Hα | 0.00† | 0.48† | 0.038 | 0.105 | -9 | 61 | 1481 | 455.75 | 16.3 |
| | | F606W: 1-comp | 1† | Hα | 0.00† | 28.91 | 0.132 | 0.095 | -29 | 73† | 1465 | 109.27 | 3.9 |
| | | F606W: 1-comp | 1† | Hα | 0.00† | 3.30† | -0.003 | 0.129 | -4 | 64 | 1484 | 586.57 | 21.0 |

## 5. DISCUSSION

Our analysis shows that to reproduce the central rotation curves of NGC 1300, a black hole of mass $6.6^{+6.3}_{-3.2} \times 10^7\ M_\odot$ is required, whilst NGC 2748 requires a black hole of mass $4.4^{+3.5}_{-3.6} \times 10^7\ M_\odot$. We use the term black hole generically to mean 'central massive object' or 'dark point mass'.

To compare our black hole measurements to those predicted by the correlations between black hole mass and bulge luminosity, we use total galaxy magnitudes (in the *J*, *K* and *H* bands) from the 2MASS extended source catalogue and allow for the disc contribution using the correction of Simien & de Vaucouleurs (1986). The total *J*, *H* and *K* magnitudes for NGC 1300 are 8.5, 7.8 and 7.6, respectively; for NGC 2748 they are 9.8, 9.1 and 8.8. The disc correction for galaxies of this type (T=4) given by Simien & de Vaucouleurs (1986) is ~ 2 magnitudes. This yields absolute *J*, *H* and *K* magnitudes of -20.9, -21.6 and -21.8 for NGC 1300, and -20.0, -20.7 and -21.0 for NGC 2748. From the correlations of Marconi & Hunt (2003) we find the predicted black hole masses to be $1.8 \times 10^7 M_\odot$ (*J*), $1.7 \times 10^7 M_\odot$ (*H*), $1.5 \times 10^7 M_\odot$ (*K*) for NGC 1300 and $7.2 \times 10^6 M_\odot$ (*J*), $6.6 \times 10^6 M_\odot$ (*H*), $7.6 \times 10^6 M_\odot$ (*K*) for NGC 2748. These estimates are in reasonable agreement with our measurements, given the r.m.s scatter of 0.3 in log $M_{BH}$ in the correlations, and our measurement uncertainties. We have obtained and are currently analysing stellar velocity dispersions for these galaxies. However, it has become clear that reliable determination of this fundamental parameter is non-trivial and the involved discussion will be postponed to a future paper (Batcheldor et al. in preparation).

We have shown that the derived masses are robust and reasonably insensitive to the fine details of the models used to reproduce the stellar luminosity density profile. The mass-to-light ratios derived in the models are reasonable for spiral galaxies (e.g. Moriondo, Giovanardi, & Hunt 1998). In NGC 1300 our models required a shift in the systemic velocity between the central and extended rotation curves of approximately 30 km s$^{-1}$. This is probably because we only observe one region of extended emission on the blueshifted side of the disc, and does not necessarily imply that there is a kinematically distinct, blueshifted central disc.

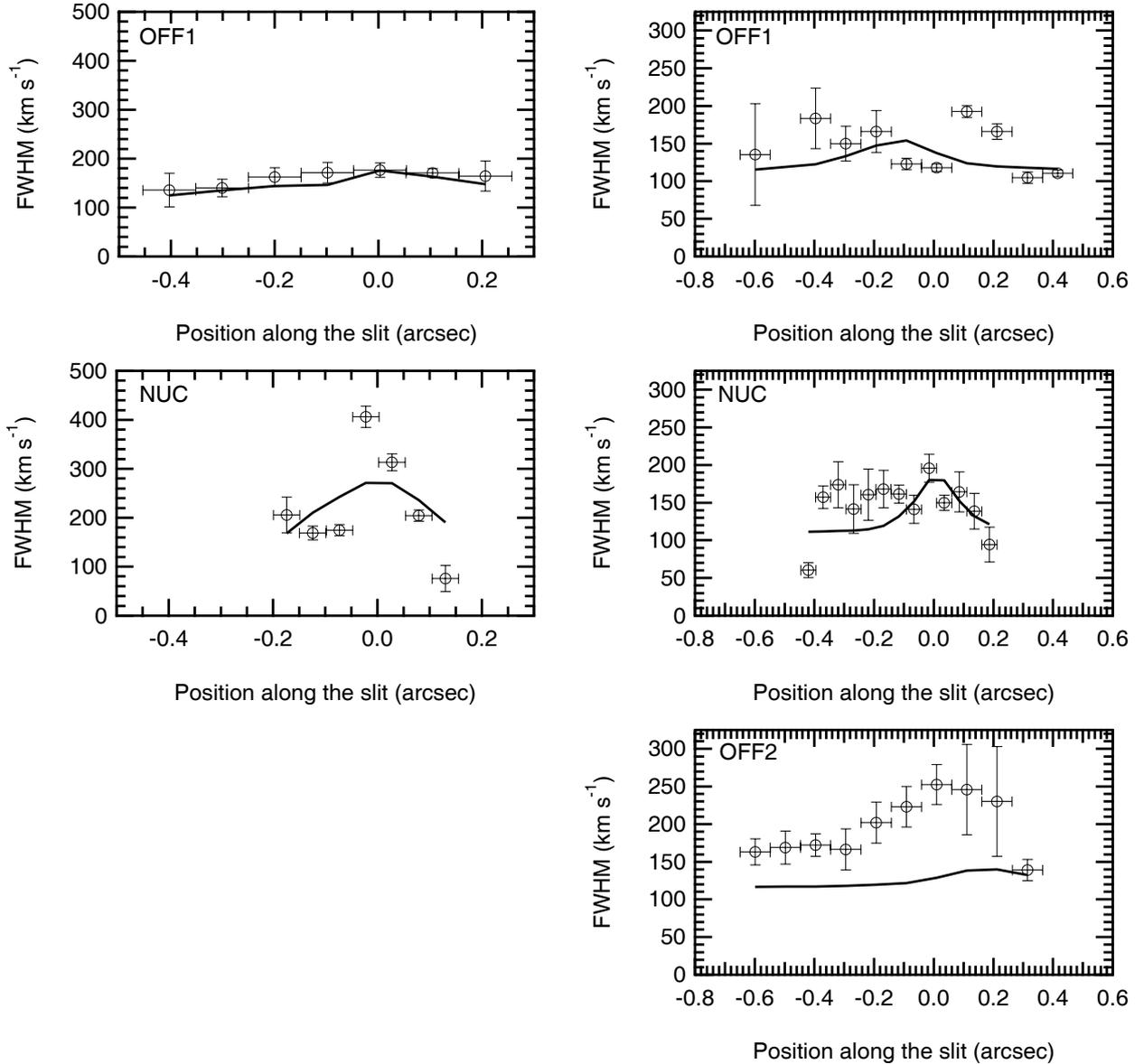

Figure 12. Central FWHM profile of NGC 1300 (left) and NGC 2748 (right) derived for the best base models using the *standard approach* (defined in section 4.4). The solid lines indicate FWHM profiles for models with intrinsic gas velocity dispersions of ~ 33 km s$^{-1}$ (NGC 1300) and ~ 37 km s$^{-1}$ (NGC 2748).

In our analysis of the rotation curves we have neglected the effect of extinction by dust. This is acceptable, as we have shown that the derived estimates of the black hole masses using *R* band and *H* band photometry (which are affected to markedly different degrees by extinction) are of the same order of magnitude. Even so we attempt to make a crude estimate of the extinction using the *H-R* colour measurements from our images. The colours are measured using apertures of 50, 100 and 500 pc and all yield similar results. This is comforting as the radial surface brightness profiles show no sign of a colour gradient – the *R* band and *H* band images have similar gradients. We assume an intrinsic *H-R* colour of -1.9 magnitudes (de Jong 1996). We measured an *H-R* colour of ~ -2.7 for NGC 1300 and ~ -3.6 for the central 50 pc of NGC 2748 (compared with ~ -3.9 for apertures of both 100 pc and 500 pc). We estimate a value for *E(H-R)* of ~ -0.8 magnitudes in NGC 1300 and a central colour excess of ~ -1.9 in NGC 2748. The extinction in the *H* band is then estimated to be 0.2 and 0.5 magnitudes, respectively, using the extinction law of Cardelli, Clayton & Mathis (1989). These extinction measurements are negligible in deriving the black hole mass if the extinction is uniform. However, they yield an increase in light of ~ 20% (NGC 1300) and ~ 58% (NGC 2748), corresponding to a decrease in the mass-to-light ratio of ~ 0.1 in log (NGC 1300) and 0.2 in log (NGC 2748). The fact that we derive similar black hole masses using bulge models based on optical and infrared images suggests that the black hole mass estimates are not sensitive to the extinction of starlight (not to say that extinction might obscure parts of the emission-line disc, which would enable better constraints in the fitting).

By relaxing the assumptions of a coplanar disc and constant stellar mass-to-light ratio we have shown that NGC 1300 still requires a black hole to reproduce the observed rotation curves. Allowing the inclination in NGC 2748 to differ in the nuclear regions does not dispose of the need of a black hole to achieve a satisfactory fit. Allowing for a different stellar mass-to-light ratio when fitting the central rotation curves (compared to that derived when fitting the extended rotation curves) weakens the need for a black hole. However, the addition of a black hole to this model again yields a better fit.

The kinematic centre of each galaxy is determined by optimisation of rotation curve models. It is of interest to locate these 'centres' on photometric images. The photometric centre of NGC 2748 is not easily identified in the optical acquisition image due to the prominent dust features. Although, these dust features are less problematic in the F160W image, there is still obscuration. We have attempted to register the two images to see if the apparent photometric centres are coincident. We find that the photometric centre of the F160W image is offset by ~ 0.2″ in the positive *y* direction and ~ 0.2″ in the positive *x* direction. These offsets correspond to roughly 4 pixels in the NUC spectra, and suggest, in fact, that the OFF1 slit lies on the nucleus of the galaxy. The *y* offset moves the centre towards the observed peak in the emission-line flux which has an offset of ~ 0.4″. However, there are certainly galaxies where the peak of emission is clearly not at the kinematic centre, e.g. NGC 2694 (see Hughes et al. 2003). Consequently, we should not be too concerned that our models have returned the kinematic centre offset from the peak of the emission-line flux. We have run additional models where the kinematic centre was fixed at the infrared photometric centre. These models yielded poor fits to the rotation curves and were not pursued. Looking at Figure 5 we see that the nuclear gas motions have both prograde and retrograde components. In the fits presented earlier, we have assumed that the prograde component, rotating with the galaxy, corresponds to the ordered motion of an emission-line gas disc. The retrograde component was discarded. However, we also investigated the case where the retrograde component was assumed to be a coplanar counter-rotating gas disc. In this case, the prograde component was assumed to be due to non-circular motions and accordingly discarded. Using the *alternative approach*, we investigated models, both with and without a black hole. We found the resultant $\chi^2$ are of a similar order to those presented in Table 5, although the OFF1 rotation curve is not well reproduced. A black hole mass of ~ $10^7$ $M_\odot$ yields the best fitting model in this approach. This shows the ambiguity that can exist even when the rotation curves are well sampled by the emission-line gas.

The presence of a black hole might be indicated by an observed rise in the central FWHM of the emission-lines. Additionally, Barth et al. (2001) suggests that an observed FWHM, that is much larger than that expected due to the instrumental broadening, may be a signature of non-circular motions. Inspection of Figure 12 shows that this is an issue in neither NGC 1300, nor NGC 2748: in both cases, the observed FWHM profile is of the same order as that which the model predicts. The model FWHM profiles displayed in Figure 12 were derived using the standard base model, but the models were minimized using both the velocity and FWHM data, simultaneously. In optimising the models (which include the instrumental broadening), the additional free parameter of the intrinsic velocity dispersion of the gas is optimised to ~ 33 km s$^{-1}$ (NGC 1300) and ~ 37 km s$^{-1}$ (NGC 2748). It is clear from the figure that although the model FWHM profiles roughly agree, in magnitude, with those observed, they do not reproduce any detail. We only observe a well defined peak in the FWHM profile of the NUC slit of NGC 1300. However, this peak is highly asymmetric and is not at all well reproduced by our model. We can conclude that while the observed FWHM profiles are consistent with the values derived in our models, they add little constraints to the derived black hole mass estimates.

**ACKNOWLEDGMENTS**

JWA acknowledges financial support from a PPARC studentship. We thank Andrew Robinson and Nial Tanvir for useful discussions. Support for proposal GO-8228 was provided by NASA through a grant from the Space Telescope Science Institute which is operated by the Association of Universities fro Research in Astronomy, Inc., under NASA contract NAS 5-26555. This publication makes use of the LEDA database.